\documentclass[conference]{IEEEtran}
\ifCLASSINFOpdf
\else
\fi
\usepackage{comment}
\usepackage[T1]{fontenc}
\usepackage[utf8]{inputenc}
\usepackage{amssymb}
\usepackage{amsmath}
\usepackage{flushend}
\usepackage{tabularx,ragged2e,booktabs,caption}
\usepackage{adjustbox,lipsum}
\newtheorem{defn}{Definition}
\newcolumntype{C}[1]{>{\Centering}m{#1}}

\begin{document}
%
\title{A New Storage Optimized Honeyword Generation Approach for Enhancing Security and Usability}

\author{\IEEEauthorblockN{Nilesh Chakraborty}
\IEEEauthorblockA{Department of Computer Science \& Engineering\\
Indian Institute of Technology Patna\\
Bihar, India - 800013\\
Email: nilesh.pcs13@iitp.ac.in}
\and
\IEEEauthorblockN{Samrat Mondal}
\IEEEauthorblockA{Department of Computer Science \& Engineering\\
Indian Institute of Technology Patna\\
Bihar, India - 800013\\
Email: samrat@iitp.ac.in}
}


%


\maketitle

\begin{abstract}
Inverting the hash values by performing brute force computation is one of the latest security threats on password based authentication technique. New technologies are being developed for brute force computation and these increase the success rate of inversion attack. Honeyword base authentication protocol can successfully mitigate this threat by making password cracking detectable. However, the existing schemes have several limitations like Multiple System Vulnerability, Weak DoS Resistivity, Storage Overhead, etc. In this paper we have proposed a new honeyword generation approach, identified as \textit{Paired Distance Protocol} (PDP) which overcomes almost all the drawbacks of previously proposed honeyword generation approaches. The comprehensive analysis shows that PDP not only attains a high detection rate of $97.23\%$ but also reduces the storage cost to a great extent.\\

\textit{Keywords $-$} Authentication; Password; Inversion attack;  Honeyword; Paired distance.
\end{abstract}


%
\IEEEpeerreviewmaketitle

\section{Introduction}
Password based authentication technique is one of the most widely used authentication technique as it nicely balances the security and usability standards. However, like any other security schemes, password based schemes have also been challenged by different attack models over times. One such recently developed attack model is \textit{inversion attack} and the model is described next.

\subsection{Inversion attack model}
\label{1-intro}
 While creating a web-account, user has to register with the site by submitting the \textit{username} and \textit{password}. System stores the \textit{username} in plain text whereas the \textit{password} is converted into hash (may be with an added salt) using hashing algorithm $\mathbb{H}$. Thus, the $i^{th}$ user's login credential $-$ stored by the system, can be represented by a tuple $<u_i, H(p_i)>$.  
Under \textit{inversion attack} model, adversary can successfully invert the hashes (evaluating $p_i$ from H($p_i$)) from the compromised password file \textit{F}. While inverting the hashes, adversary first derives a password string using some existing techniques \cite{john}, \cite{weir}, \cite{ma}. Then adversary matches the password string (appending the salt, if required) after converting it into hash value using $\mathbb{H}$. If the obtained hash value gets matched with stored hash value then adversary becomes successful in inverting the hashes.

Initially, brute force attack was conducted by guessing many possible combinations to break a password. But time complexity using this approach used to be very high as attacker tries for every possible options for cracking a password. One of the publicly available password cracking algorithm which significantly reduces the time complexity of \textit{inversion attack} $-$ was proposed by John and Ripper in $2008$ \cite{john}. In $2009$, based on the concept of probabilistic context free grammar, Weir et.al. were able to crack $28\%-129\%$ more passwords than Jhon and Ripper password cracking technique. Recently proposed technique by Ma.et.al \cite{ma} $-$ uses Markov chain model for password cracking and shows significant improvement over proposed algorithm by Wier et.al.

\textbf{\textit{Evidences :}} There are some \textit{strong evidences} of \textit{inversion attack} which threats security of some reputed web based organizations. In recent past, almost 50 millions passwords of Evernote have been compromised \cite{evernote} by performing \textit{inversion attack}. Giant web-service based organizations like LinkedIn, Yahoo, RockYou have gone through the same misery \cite{linkedin}. So there is an urgency for developing an improved \textit{honeyword} based framework, robustly handles the \textit{inversion attack}. 

\subsection{Existing security techniques}
\label{2-intro}
 Few security techniques have been developed to address this security issue. There are some tricks using which user's password can be transformed into some hash value which is harder to invert. This type of login set up increases the login time and does not make successful password cracking detectable \cite{hard-inversion}. Another alternative may be $-$ setting up few fake login accounts by the administrator. An adversary, who successfully inverts the hash value of any such account, system detects the security breach. But with some careful analysis, adversary can distinguish the real \textit{usernames} from the system generated \textit{usernames} \cite{fakeaccount2}.     

\textit{Honeyword} based approaches have shown some significant potential while providing security against \textit{inversion attack}. Using this approach system maintains a list of passwords which contains the real user's password along with some system generated passwords, known as \textit{honeywords}. System generates these \textit{honeywords} by using any of the underlying \textit{honeyword} generation algorithms such as - \textit{take-a-tail} \cite{juels}, \textit{modelling-syntax} \cite{kamouflage} etc. Once password file \textit{F} is compromised and adversary enters any of the \textit{honeywords} from the password list of $W_{i}$, system identifies the attack and takes necessary actions depending upon the security policy.

\subsection{Motivation and Contribution}
\label{mc}
Among all the \textit{honeyword} generation techniques proposed so far $-$ \textit{take-a-tail} approach (see details in Section \ref{o-l-t}) sets strongest security standard among all \cite{juels}. But the technique threats usability standard to a great extent as user with \textit{n} different login accounts, has to remember \textit{n} different system generated information as a part of his login credential. Moreover, we have found that all proposed \textit{honeyword} generation techniques require to store \textit{k-1} ($k > 1$) \textit{honeywords} to lure the attackers. Storing \textit{k-1} extra information for each \textit{username}, magnifies the storage cost to a great extent.   

Thus, from the existing literature surveys, we gist the motivations behind this work and those are summarized as below $-$
\begin{itemize}
\item \textbf{Motivation 1 :} All existing \textit{honeyword} generation techniques store \textit{k-1} decoy passwords to detect the security breach. Thus, the storage cost is required to be minimized which increases with the number of users.\\
\item \textbf{Motivation 2 :} Till date, though ``take-a-tail" sets highest security standard but the method threats the usability standard badly as user requires to remember \textit{n} different system generated information for \textit{n} different accounts. Like in \cite{erguler-first}, we also feel that remembering \textit{n} different informations for \textit{n} accounts is infeasible for most of the users due to limitation of human memory \cite{limitation}. Thus, a \textit{honeyword} based security architecture is needed to be developed which ensures same security standard as ``take-a-tail" by enhancing the usability standard.  
\end{itemize}

Motivated by the above mentioned facts we have made following two major contributions in this paper $-$
\begin{itemize}
\item \textbf{Contribution 1 :} We propose a new method termed as \textit{Paired distance protocol} (PDP) for generating \textit{honeywords}. The method stores only a single information to generate the \textit{honeywords} and thus, minimizes the storage burden significantly.\\

\item \textbf{Contribution 2 :} Using the proposed technique, users require to remember only a single information (of their own choice) to maintain \textit{n} different accounts . Thus, instead of remembering \textit{n} system generated different information for \textit{n} different accounts (like in ``take-a-tail"), user only remembers a single information and still can avail the same security standard as ``take-a-tail".   
\end{itemize}  


\textbf{Roadmap $-$}  In Section \ref{overview} we give an overview and limitations of existing \textit{honeywords} generation algorithms. In Section \ref{proposed} we introduce the proposed PDP approach and show how proposed scheme works to detect the attack? Security and usability analysis of PDP is illustrated in Section \ref{sec} and Section \ref{use} respectively. In Section \ref{store} we show how PDP minimizes the memory overhead? A detailed comparative analysis of PDP with existing security techniques is provided in Section \ref{compare}. In Section \ref{relate} we give a brief outline of existing work in this direction. Finally we conclude and give some future directions of our work in Section \ref{conclusion}.   

\section{An overview on honeyword based authentication technique and it's limitations}
\label{overview}
In this section, first we describe the working principal of \textit{honeyword} based authentication scheme. There after we present limitations of existing schemes, proposed in this direction. But prior to that some of the related notations that we are going to use, are presented in Table \ref{notation}. 

\begin{table}[!h]
\centering
\begin{tabular}{ |c | c | }
\hline
\textbf{Notations} & \textbf{Meaning} \\
\hline
  $u_{i}$ & $i^{th}$ user in system \\ 
\hline  
  $p_{i}$ & password of $i^{th}$ user  \\
\hline
$W_{i}$ & tuple of passwords stored for $u_{i}$ \\
\hline  
$k$ & number of elements in $W_{i}$\\
\hline  
$c_{i}$ & index of correct password in $W_{i}$ \\
\hline  
\textit{sweetword} & each element of $W_{i}$\\
\hline
\end{tabular}
\caption{Related notations}
\label{notation}
\end{table}

\subsection{Honeyword based authentication technique}

As mentioned in Section \ref{2-intro}, a \textit{honeyword} generation scheme maintains a list $W_{i}$ against each \textit{username} $u_{i}$. The index of correct password is maintained in another file in a different system (known as ``honeychecker"). The basic idea is $-$ even if $W_{i}$ is compromised and adversary successfully inverts each \textit{sweetword} then also adversary gets confused about original password of user as user's complete password information is distributed over two different systems. If adversary picks any \textit{sweetword} from the $I_{i}^{th}$ index of list $W_{i}$ and submits that against user id $u_{i}$ then index of that \textit{sweetword} ($I_{i}$) is directed to the ``honeychecker". If $I_{i}$ gets matched with $c_{i}$, honey-checker  directs a positive feedback to the system administrator otherwise, ``honeychecker" directs a negative feedback. Depending upon the security policy, system administrator takes necessary actions according to the received feedback from the ``honeychecker". Thus,\textit{honeyword} based system provides distributed security which is harder to compromise as a whole \cite{juels}.

\subsection{Limitations of honeyword based authentication technique}
\label{limitations}
Though existing \textit{honeyword} based approaches can provide security against brute force attack but they have few limitations. The limitations are described below $-$

\textbf{(a) Storage overhead $-$} Using \textit{honeyword} generation approach, system needs to store \textit{k-1} more passwords for each user account. Thus for a system storing \textit{n} users accounts, needs to store \textit{n} $\times$ \textit{(k-1)} extra information which magnifies the storage cost to a great extent. This is identified as one of the major drawback of any \textit{honeyword} generation approach.\\

\textbf{(b)} \textbf{Co-relational hazard} $-$ If there exists a relationship between \textit{username} and the \textit{password} (e.g. \textit{username} as \textit{football} and \textit{password} as \textit{maradona}) then the original password of user can easily be identified from the list of $W_{i}$. In such cases \textit{honeywords} can not mask the original password.\\

\textbf{(c)} \textbf{Distinguishable well-known password patterns} $-$ If user uses a password which is related to some well known object/fact, then attacker can easily identify the original password. For example, some of the passwords belong to this category are $-$ \textit{bond007}, \textit{james007}, \textit{007bond} and \textit{007007} and were found from the list of $10000$ most common passwords \cite{weakpassword01}.\\

\textbf{(d)} \textbf{Issue related to DoS resistivity} $-$ If adversary can guess the \textit{honeywords} while he knows the original password of user, then adversary can intentionally submit \textit{honeyword} to generate a \textit{false} negetive feedback signal by the ``honeychecker" (while \textit{F} is not compromised). Adversary can submit \textit{honeywords} from many user accounts (either by creating them or, by knowing original password of users by shoulder surfing attack \cite{kwon2014}) so that system understands the password file \textit{F} has been compromised when it is actually not. If system senses submission of \textit{honeywords} from too many accounts then system may block the whole web server. This is known as Denial-of-Service (DoS) attack \cite{dos}. Thus, original password of user must not give any idea about system generated \textit{honeywords} to avoid DoS attack. Some of the \textit{honeyword} generation techniques like $-$ \textit{chaffing-by-tweaking digits} \cite{juels} provide weak security against such kind of attack while some others like $-$ \textit{modelling-syntax-approach} \cite{kamouflage} provide strong security against DoS.\\      

\textbf{(e)} \textbf{Issue related to Multiple System Vulnerability} $-$ If a user uses same password in two (or more) different systems (where systems are \textit{using same honeyword generation algorithm}) and an adversary gets access to both the systems, then Multiple System Vulnerability may occur. In this case, adversary may obtain obtains two lists of $W_{i}$ for user $u_{i}$. Let $W_{i}^{S_j}$ denotes list of \textit{sweetwords} for user $u_{i}$ in the system $S_{j}$. Now if generated \textit{honeywords} belong to $W_{i}^{S_p}$ and $W_{i}^{S_q}$ (where p $\neq$ q) are different (probability of which is close to $1$) then by performing intersection operation $W_{i}^{S_p} \cap W_{i}^{S_q}$ adversary obtains the original password. This is identified as Multiple System Vulnerability (MSV) of \textit{honeyword} based authentication technique.\\ 

\textbf{(f)} \textbf{Issue regarding Typo safety} $-$ A \textit{honeyword} generation technique is called \textit{typo safe} if typing mistake of users during entering of the password does not get match with any of the \textit{honeywords}. ``Chaffing-by-tweaking" \cite{juels} methods are not much \textit{typo safe} as a legitimate user may accidentally submit a \textit{honeyword}. Consider the following example where user chooses his password as \textit{road8}. Now ``chaffing-by-tweaking-digits" may produce following list of \textit{sweetwords} for \textit{k = 6}

\begin{center}
\begin{tabular}{c c c c c c}
\textit{road9} & \textit{road2} & \textit{road5} & \textbf{road8} & \textit{road4} & \textit{road6}
\end{tabular}
\end{center}
   
From the above list of \textit{sweetwords} it can be seen that, \textit{honeywords} are created from user password \textit{road8} by replacing the single digit. So the probability that user typing mistake will match with a \textit{honeyword} for \textit{k = 6} is $5/9$ (if user mistakenly enters a wrong digit while prefix of the password (here \textit{road}) remains same). 

``Take-a-tail" method proposed by Juels and Rivest, successfully addresses all the above mentioned drawbacks except \textit{storage overhead}. A brief overview and limitation of ``take-a-tail" is presented next before we go into details of the proposed \textit{PDP} protocol. 

\subsection{Take-a-tail : An overview and limitation}
\label{o-l-t}  
Using ``take-a-tail" user first provides the \textit{username} and his \textit{password} choice during the course of registration to a system. System then generates a random string of length $\ell (> 0)$ consisting of alphabets and (or) digits, identified as \textit{tail} in \cite{juels}. While login to the system, user requires to submit \textit{username} and \textit{password} along with the system generated tail. Thus, the user registration interface can be described as shown in Fig. \ref{interface}.

\begin{table}
\centering
\begin{tabular}{|c c|}
\hline
Enter Username : & Alice\\
Enter Password Choice : & $******$\\
\textbf{\textit{Append $613$ to complete your password}} & \\
Enter Revised Password : & $*********$\\
\hline
\end{tabular}
\vspace{0.3cm}
\captionof{figure}{Registration interface of take-a-tail}
\label{interface}
\end{table}

The example in Fig.\ref{interface} shows that system generates $613$ as a \textit{tail}. During each login session, user requires to submit the \textit{password} along with the appended \textit{tail}. System generates the \textit{sweetwords} by the ``chaffing-by-tweaking" tail. Thus, for the password \textit{street$613$} ($613$ is tail here) the following probable list of \textit{sweetwords} for \textit{k = 5} is $-$      

\begin{center}
\begin{tabular}{c c c c c c}
\textit{street124} & \textit{street498} & \textit{street668} & \textbf{street613} & \textit{street153} 
\end{tabular}
\end{center}
Thus, even if there exist a \textit{co-relational hazard}, adversary can hardly distinguish the user's original password from the list of \textit{sweetwords}. As tails differ for each login account of a user thus, MSV is also avoidable even if user chooses the same password. Knowing the original password (along with \textit{tail}) also makes it difficult for the adversary to guess the \textit{honeywords} and as a result provides a standard security against DoS attack. Thus, with some careful observation it is easily understandable that ``take-a-tail" overcomes all the drawbacks mentioned in Section \ref{limitations}, except the \textit{storage overhead}.   

As discussed earlier, the \textit{limitation} of this approach remains in remembering \textit{n} different system generated \textit{tails} for \textit{n} different login accounts and this reduces the usability factor to a great extent. Other than this, ``take-a-tail" imposes same storage overhead as other \textit{honeyword} generation approaches. Next we introduce the proposed \textit{PDP} $-$ a storage optimized \textit{honeyword} generation approach with enhanced usability factor.

\section{Proposed methodology}
\label{proposed}
Our proposed approach is identified as \textit{Paired Distance Protocol} or PDP. Using the proposed approach user needs to provide three information $-$ (a) Username (b) Password and (c) a Random String \textit{RS} of length $\ell$ $-$ containing alphabet and numbers of user's own choice. The default length of \textit{RS} is set as $3$. Thus, along with \textit{password}, user has to remember another secret information as \textit{RS}. Initially, it may appear to be an overhead but \textit{RS} provides several advantages which we have discussed elaborately in the subsequent sections. Few important characteristics of \textit{RS} are discussed next.

 Using PDP user can use the same \textit{RS} for different systems. However, users are strongly advised to choose a random \textit{RS} (e.g. not a dictionary word). If chosen \textit{RS} by user is hard to guess and doesn't follow either a specific pattern (e.g. sequential keystroke) or, dictionary word (e.g. fox) and there is no correlation with either \textit{username} or \textit{password} (e.g. \textit{username} $-$ jerry, \textit{password} $-$ face and \textit{RS} $-$ eye) $-$ then \textit{randomness} of \textit{RS} is considered as high. No element in \textit{RS} should get repeated to avoid DoS attack (see detail in Section \ref{dos}). 

Our assumption behind setting up high \textit{randomness} of string \textit{RS} by user is valid because remembering \textit{RS} doesn't impose much overhead on users as $-$
\begin{itemize}
\item The string length of \textit{RS} is less (considered as $3$ to avoid specific pattern like \textit{date of birth}).
\item User may use same \textit{RS} for different login accounts.
\end{itemize} 

The registration interface using \textit{PDP} can be described by the Fig. \ref{interface-01}

\begin{table}[!ht]
\centering
\begin{tabular}{|c c|}
\hline
Enter Username : & Alice\\
Enter Password Choice : & $******$\\
\textbf{\textit{Choose a random string}}  & \\
\textbf{\textit{to complete your password}} & \\
Enter Revised Password : & $*********$\\
\hline
\end{tabular}
\vspace{0.3cm}
\captionof{figure}{Registration interface of PDP}
\label{interface-01}
\end{table}

The \textit{fundamental differences} between ``take-a-tail" and proposed approach $-$ from usability perspective, is presented in Table \ref{differences}. 
\begin{table}[!ht]
\centering
\begin{tabular}{|c|c|}
\hline
Take-a-tail & PDP\\
\hline
User remembers           & User remembers  \\

the extra information    & the extra information of \\

\textbf{generated by system} & \textbf{his own choice}\\
\hline
For \textit{n} different   & For \textit{n} different  \\
accounts, user must        & accounts, user may        \\ 
remember \textbf{\textit{n}} information   & remember \textbf{single} information\\
\hline 
\end{tabular}
\vspace{0.2cm}
\caption{Differences between ``take-a-tail" and PDP from usability perspective}
\label{differences}
\end{table}

Next we elaborate on how \textit{honeyword} can be generated by using our proposed approach?  

\subsection{Setting up the honey circular list}
\label{su-hcl}
\begin{figure}
\centering
\includegraphics[width=0.30\textwidth]{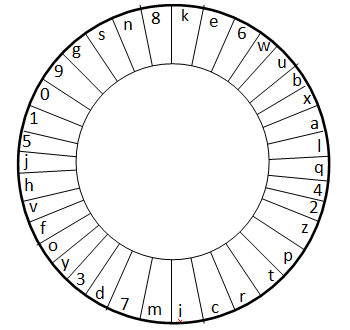}
\caption{Honey Circular List : Contains alphabets and digits in random order}
\label{hcl}
\end{figure}

Recently proposed  Sauth approach \cite{sauth} shows how different web-servers can collaborate to achieve a high security standard. We also follow the same principle to get rid of \textit{MSV} issue of \textit{honeyword} based authentication scheme. First of all a circular list $-$ identified as \textit{honey circular list} or \textit{hcl} of length $|\text{hcl}|$ is created which holds the alphabet and digits in random order. The default value of $|\text{hcl}|$ is considered as $36$ here. For default value of $|\text{hcl}|$  we show one instance of \textit{hcl} in Fig. \ref{hcl}.

This \textit{hcl} is then securely distributed to \textit{m} different systems (may be \textit{facebook}, \textit{google} etc.), participating in creating \textit{honeywords} using PDP protocol. The \textit{hcl} is maintained in the password file \textit{F}. The utility of \textit{hcl} for avoiding DoS is elaborated in Section \ref{dos}.

\subsection{Maintaining the user database}
\label{mud}
While maintaining the user's login information in the database, system does the following tricks. System first stores \textit{username} along with the \textit{password} (may be in the hash format) of user. System then measures the distance between the consecutive elements of \textit{RS} with respect to the elements stored in \textit{hcl}. The distance between any two elements of the \textit{hcl} is known as \textit{paired distance} and is defined as follows $-$

\begin{defn}
\textbf{Paired distance:} Paired distance between two elements $e_1$ and $e_2$, denoted as Pr($e_1,e_2$) $-$ is the number of cells that has to be traversed in clock wise direction in the \textit{honey circular list} to reach from element $e_1$ to element $e_2$.
\end{defn}
  
Suppose user chooses \textit{RS} as \textit{``tp7"} then paired distance can be calculated as Pr(t,p) $=$ 35 and Pr(p,7) $=$ 6. Along with \textit{username} and \textit{password}, system stores this paired distances between two consecutive elements of \textit{RS} separated by ``$-$" (e.g. $35-6$).

\begin{defn}
\textbf{Distance chain :} Distance chain is the set of \textit{n-1} paired distances (separated by $``-"$) between every two consecutive elements of \textit{RS}, having length \textit{n}. 
\end{defn}

Along with \textit{username} and \textit{password}, (instead of storing \textit{k-1} \textit{honeywords}) system maintains the \textit{distance chain} derived from \textit{RS} in password file \textit{F}. While analysing, we have found a special property of \textit{distance chain} and we identified it as \textit{uniqueness property} which is defined next.

\textit{\textbf{Uniqueness property of distance chain :}} Given a \textit{hcl} and a particular distance chain $-$ \textit{RS} can be uniquely derived if first element of \textit{RS} is known.

Let's describe this with the previous example. Suppose \textit{distance chain} $35-6$ is known along with the first element of \textit{RS} which is \textit{``t"}. Now with respect to the \textit{hcl} shown in Fig. \ref{hcl}, string \textit{``tp7"} can be derived by performing reverse calculation which is unique. 
Now if first element of \textit{RS} is unknown then starting with each element of \textit{hcl} a given \textit{distance chain} can be derived. For example, if \textit{distance chain} is $35-6$ then by reverse calculation, string \textit{``k8b"}, \textit{``ekx"} etc can be derived using the \textit{hcl} shown in Fig. \ref{hcl}. Thus for a given \textit{distance chain} total number of possible \textit{RS} is $|\text{hcl}|$.

\textbf{\textit{Necessity of choosing RS :}} The strength of  PDP depends on \textit{randomness} of string \textit{RS}. The \textit{honeywords} are generated using the string \textit{RS} and \textit{hcl}. While analysing, we found that normal tendency of users is to use meaningful phrase in their password \cite{weakpassword01} (e.g. \textit{secret123}, where \textit{secret} is meaningful). From the \textit{distance chain} stored by the system, adversary becomes able to derive different possible strings, which also contain the chosen \textit{RS} by user. If user password is used in place of \textit{RS} for generating the \textit{distance chain} then adversary may able to distinguish user's original password. The reason behind this is $-$ due to random organization of characters in \textit{hcl}, the probability of deriving a string $-$ containing a meaningful phrase (except the original password of user), is very less. Now as meaningful phrase is used in most of the passwords, created by user thus, from the derived strings from the \textit{hcl} and the \textit{distance chain}, adversary can easily distinguish user's original password from the non-meaningful derived strings. Hence there is a necessity of choosing \textit{RS} which is random enough and can't be easily guessed by the adversary.

\subsection{Maintaining the honeychecker} In existing approaches, generally \textit{honey-checker} maintains the index of original password of the user along with \textit{username}. In the proposed approach, along with \textit{username}, ``honeychecker" maintains first character of the \textit{RS}, chosen by the user.    

\subsection{Working principal of the proposed scheme}
During login, when user submits the login credentials, system first checks the correctness of the password entered by the user. If the password entered by the user is incorrect then system straightaway denies the user login.

 If password entered by the user is correct then system derives the \textit{distance chain} from the submitted \textit{RS}. If derived \textit{distance chain} doesn't get matched with the stored \textit{distance chain} in the password file \textit{F} then system denies the user.
 
  If derived \textit{distance chain} gets matched with stored one, system then communicates the first element of \textit{RS} (submitted by the user) to the ``honeychecker". If first element of \textit{RS} submitted by the user gets matched with the element stored in ``honeychecker" then ``honeychecker" directs \textit{positive feedback} to the administrator otherwise ``honeychecker" directs a \textit{negative feedback} by detecting the attack.    
  
\subsection{Evaluating the probability of detecting the attack}
Instead of storing \textit{k-1} extra information PDP is just storing one extra information as \textit{distance chain}. Now there can be $|\text{hcl}|$ number of probable \textit{RS} corresponding to a \textit{distance chain}. Thus, by storing a single information, systen confuses the attacker among $|\text{hcl}|$ different possibilities. For default value of $|\text{hcl}| = 36$, the attack can be detected with the probability of $35/36$ (or, $97.23\%$ chances).

\subsection{Password meter}
\label{pass-meter}
Password meter shows how random \textit{RS} is? If \textit{randomness} of the \textit{RS} is high then password meter shows strong signal otherwise, it shows weak signal. Below we show some of the instances of choice of \textit{RS} for which \textit{randomness} is low $-$
\begin{itemize}
\item \textit{RS} is concatenated with user \textit{password} and if it makes some dictionary word (e.g. \textit{password} $-$ rab, \textit{RS} $-$ bit).
\item If \textit{RS} itself is a dictionary word (e.g. fox).
\item \textit{RS} follows a specific pattern (like, sequential keystroke), distinguishable by attacker.
\end{itemize} 
Users are recommended to change their \textit{RS} if password meter shows low \textit{randomness}. If there exist a co-relation among \textit{username}, \textit{password} and \textit{RS} then also \textit{randomness} of \textit{RS} becomes low, though password meter is not able to address that.

\section{Security standards}
\label{sec}
There are three well defined security parameters for evaluating the robustness of any \textit{honeyword} generation algorithms $-$ \textit{(a) Flatness (b) DoS resiliency and (c) Security against MSV}. 
Next we will evaluate the strength of PDP by considering these three security standards along with a new security factor termed as \textit{collaborative security}.

\subsection{Flatness}
\label{flat}
 If system maintains \textit{k} sweetwords against a user $u_i$ then attacker may get confused among \textit{k} possible options once $W_{i}$ is compromised. Now sometimes it may happen that, adversary can easily identify the password chosen by the user from the list $W_{i}$ (e.g. if there exists a correlation between \textit{username} and \textit{password}). A \textit{honeyword} generation algorithm is said to be \textit{perfectly-flat} if adversary has no advantage while identifying the user's original password from the list of $W_{i}$. If the \textit{honeyword} generation algorithm is \textit{perfectly-flat} then probability of selecting the original password of user from list $W_{i}$ is $1/k$. If the probability of selecting user password from the list $W_{i}$ is slightly greater than $1/k$, then the \textit{honeyword} generation algorithm is called \textit{approximately-flat}. A good \textit{honeyword} generation algorithm is required to be \textit{perfectly-flat}.

PDP becomes a \textit{perfectly-flat} technique if the randomness of chosen \textit{RS} is high. 

\subsection{DoS resiliency}
\label{dos}
 Performing DoS attack (discussed in Section \ref{limitations}) is highly impossible on a \textit{PDP secure system}. DoS attack is only possible if adversary can generate a \textit{distance chain} that is maintained by the system for any different \textit{RS} not chosen by user. As \textit{RS} not allows repetition of characters thus, adversary requires the knowledge of orientation of characters in the \textit{hcl} to perform the attack. For example, if \textit{RS} allows repetition of characters then adversary may create a \textit{distance chain} made from characters \textit{RRR} and while login, adversary may submit \textit{RS} as \textit{SSS} to perform DoS attack. This is because both the \textit{RS} derive same \textit{distance chain} as $0-0$ but first character stored in ``honeychecker" (here \textit{R}) mismatches with first character of submitted \textit{RS} (here \textit{S}). Hence adversary becomes successful to accomplish the DoS attack.
 
As all the elements in \textit{RS} get differ from each other thus, without knowing the orientation of characters in the \textit{hcl}, the probability of generating a given \textit{distance chain} by submitting a \textit{RS} (which is not chosen by user) can be calculated by Equation \ref{dos-eqn}.

\begin{equation}
|hcl|-1 \times {\sum_{i=0}^{\ell - 1} {\frac{1}{|hcl| - i}}}
\label{dos-eqn}
\end{equation}

For the default values of parameters ($\ell = 3$ and $|\text{hcl}| = 36$) the probability of successful DoS attack becomes $0.81 \times 10^{-3}$, which is very less.
\subsection{Security against MSV}
\label{msv}
A user may use \textit{same password} in $\textit{Z}$ ($> 1$) different systems which use the same \textit{honeyword} generation algorithm. Now if two such different systems are compromised then adversary can get the original password of user by performing an intersection operation. This is because, for a given password, a \textit{honeyword} generation algorithm produces different \textit{honeywords} at each run with very high probability (close to 1) \cite{juels} \cite{kamouflage}. Thus, for a given password, a \textit{honeyword} generation algorithm produces different \textit{honeywords} for each system.

Now if PDP is adopted by $\textit{Z}$ different systems which secretly share a \textit{hcl}, then for a given \textit{RS} all the system generated \textit{distance chains} will be same. Thus, even if two different accounts of a user (using the same password) are compromised $-$ then also MSV will not occur.

Another important observation is that, by identifying user $u_i$'s login credentials in a system, adversary would not be able to guess the password or \textit{RS} used by $u_i$ in other accounts unless both \textit{password} and \textit{RS} are same. 

\subsection{Collaborative security}
\begin{figure}
\centering
\includegraphics[width=0.45\textwidth]{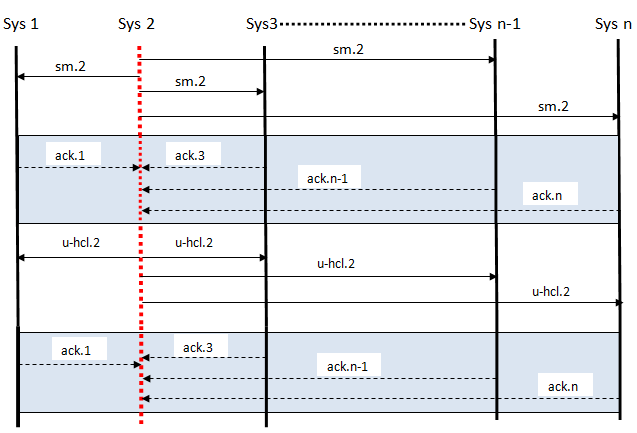}
\caption{Above figure shows PDP protocol is used by \textit{n} different systems by using same \textit{hcl}. \textit{Vertical dotted line} indicates the compromised system. \textit{sm.i} indicates security message from $i^{th}$ system. \textit{ack.i} denotes the acknowledgement generated by $i^{th}$ system. \textit{u-hcl.i} indicates the updated \textit{hcl} generated by $i^{th}$ system. The compromised system generates updated \textit{hcl} after it receives acknowledgement from all other \textit{n-1} systems.}
\label{up-hcl}
\end{figure}

In PDP approach, if a system senses that the \textit{hcl} has been compromised (after ``honeychecker" generates \textit{negative feedback} for $\mathbb{E} (> 1)$ users accounts) then it will broadcast a security message (\textit{sm}) to all other systems $-$  generating \textit{honeywords}, by using the same \textit{hcl}. Once such \textit{sm} is received by all the systems, a new \textit{hcl} is being generated (with a different orientation of same set of characters) by the compromised system and is received by all the systems under this PDP approach. In Fig. \ref{up-hcl} we give a pictorial overview on how \textit{hcl} is being shared among different systems?

 After receiving the new \textit{hcl}, each system does the following things $-$
\begin{itemize}
\item From the first character of \textit{RS} stored in the ``honeychecker" and with the reference of \textit{distance chain} and previous \textit{hcl}, system generates (and temporary stores) \textit{RS} for each user. For example, after receiving \textit{t} as the first character of \textit{RS} from ``honeychecker" system can derive the complete \textit{RS} as \textit{tp7} with the help of \textit{distance chain} $35-6$ and with the \textit{hcl} shown in Fig. \ref{hcl}. 
\item After deriving the string \textit{RS} for each user, the previous \textit{hcl} is replaced by new \textit{hcl} with the different orientation of characters.
\item By calculating the paired distance between the consecutive characters of \textit{RS} from the new \textit{hcl}, system derives the new \textit{distance chain} for each user and communicates the first character of \textit{RS} to ``honeychecker".
\item After setting the complete password information, system then removes the stored \textit{RS} for each user.
\end{itemize}

Thus, by setting the high \textit{randomness} of string \textit{RS}, user may set a high security standard in terms of flatness, DoS resistivity and security against MSV.

\begin{table*}
\centering
\resizebox{1\textwidth}{!}{
\begin{tabular}{|c||c|c|c||c|c|c|c|}
\hline
Honeyword &  Flatness                  & DoS           &   Security     &  System         & Typo    & Stress on    & Storage\\
 method   &                            & Resiliency    & against MSV    &  interference   & safety  & memorability &overhead\\
\hline
CTD       &   $1/k$  if U $\approx$ G  &  low          &     low        &    no           & low      &      low    & k-1  \\
\hline
modelling
-syntax   &   $1/k$  if U $\approx$ G  &  high         &     low        &    no           & high     &      low    & k-1 \\ 
\hline
take-a-
tail      &    $1/k$ (unconditionally) &  low          &     high       &    high         & high     &      high   & k-1 \\
\hline
PDP       &    $1/k$ $\circledast$     &  high         &     high       &    low          & high     &      low    & 1 \\
\hline
\end{tabular}}
\vspace{0.5cm}
\caption{Comparative usability analysis of \textit{honeyword} generation methods. U $\approx$ G indicates, if \textit{honeywords} are distributed like user chosen password from the adversary point of view. $\circledast$ indicates if randomness of \textit{RS} is high. Storage overhead shows the extra information system has to store in password file \textit{F}.}
\label{comp}
\vspace{0.5cm}
\end{table*}   
   
\section{Usability standards} 
\label{use}
The usability standard, set by a \textit{honeyword} generation approach can be measured in terms of three parameters $-$ \textit{(a) Typo safety} \textit{(b) System interference} and \textit{(c) Stress on memorability}. Each of these are discussed next.

\subsection{Typo safety}
A \textit{honeyword} generation algorithm is called typo safe if typing mistake of users doesn't lead to generate a \textit{negative feedback} signal by \textit{honey-checker}. Using PDP, \textit{honey-checker} generates a \textit{negative feedback} signal only if the string other than \textit{RS} derives a \textit{distance chain} that gets matched with the stored \textit{distance chain}. 
While typing the \textit{RS}, user can enter either \textit{(a)} sub part of \textit{RS} as wrong or, \textit{(b)} all the elements of \textit{RS} as wrong. If user enters sub part of \textit{RS} as wrong (e.g. instead of \textit{tp7}, if he enters \textit{tp8}) then it will never evaluate a \textit{distance chain} which gets matched with the stored one. If user enters all the elements of \textit{RS} wrong (which may rarely happen) by typing mistake, the probability that a same \textit{distance chain} (liked stored one) will be generated, can be derived by Equation \ref{typo-eqn}, same as Equation \ref{dos-eqn}.

\begin{equation}
\text{Prob} = |hcl|-1 \times {\sum_{i=0}^{\ell - 1} {\frac{1}{|hcl - i|}}}
\label{typo-eqn}
\end{equation}  

For the default values of parameters Equation \ref{typo-eqn} can be evaluated as $0.81 \times 10^{-3}$ which is significantly less. Thus, PDP is highly typo safe.

\subsection{System interference}

System interference of a \textit{honeyword} system reflects $-$ how much a \textit{honeyword} generation algorithm influences the password choice of the user? If user needs to adjust his password according to the \textit{honeyword} generation policy of the system then there exists system interference. 
Here we define three level system interference $-$ a) \textit{High system interference $-$} where user needs to adjust/manipulate his password choice based on the parameter value provided by system. Like in ``take-a-tail" \cite{juels}, choosing a three digit tail may be considered as a parameter whereas ``635" may be considered a valid parameter value set by the system. This value is needed to be remembered by the user. b) \textit{Low system interference $-$} where user sets the value of the parameter to manipulate his password choice. The proposed method PDP, requires to set the \textit{RS} (considered as a parameter) where the parameter value (i.e. elements of \textit{RS}) is chosen by the user. We consider this as \textit{low system interference} as system gives the opportunity to user to choose the value of his own choice. c) \textit{No system interference $-$} where user needs not to manipulate his password choice.

\subsection{Stress on memorability}

There exists a relation between system interference and stress on memorability. If system interference of a \textit{honeyword} scheme is high, then user has to remember different system generated information for different login accounts. These increase the stress on memorability. On the other hand, using a \textit{honeyword} generation scheme having low/no system interference, user may use the same login credential for different login accounts. Thus, stress on memorability becomes low in this case. Proposed PDP approach imposes low stress on memorability because of it's low system interference.

\section{Storage cost}
\label{store}
Using the previously proposed \textit{honeyword} generation algorithms system maintains \textit{k-1} extra passwords along with the original password of user, in the password file \textit{F}. On the other hand index of the original password of the user is maintained in ``honeychecker" server. If we assume that for storing a single password, system requires $\theta$ memory space then for storing password information of \textit{n} users, would require $n  \theta k$ space. Whereas the required space in ``honeychecker" is $n  \theta$. 
Using PDP, for each user, system stores two information (password and \textit{distance chain}). Thus, the password storing cost for \textit{n} users in password file \textit{F} becomes $2 n \theta$. Though, PDP maintains a \textit{hcl} of size  $|\text{hcl}| \theta$ $-$ but the storage cost does not depend on number of users. So storage cost of \textit{hcl} is negligible.  As memory cost of storing an index value in the ``honeychecker" is very similar with storing a digit/alphabet so, required space in ``honeychecker" is same as $n  \theta$. Thus, PDP saves a memory overhead $\textit{n} \theta \textit{(k-2)}$ . 

As any standard \textit{honeyword} system maintains the value of k as $20$ for moderate detection rate \cite{juels} thus PDP saves a memory overhead $18\textit{n}\theta$ which is a huge benefit.

\section{Comparative analysis}
\label{compare}
In the scope of this section, we compare PDP with some of the recently proposed \textit{honeyword} generation approaches \textit{(a) Chaffing-by-tweaking-digits} (CTD) \cite{juels}, \textit{(b) Take-a-tail} \cite{juels} and \textit{(c) Modelling-syntax-approach} \cite{kamouflage}, in terms of security and usability standards (shown in Table \ref{comp}).

Above table shows that by choosing a high random \textit{RS}, user can avail the same security standard as ``take-a-tail" in terms of providing security with respect to \textit{Flatness} and \textit{MSV}. The limited strength of ``take-a-tail" in term of providing security against DoS attack \cite{juels} has also been overcome by PDP approach. Thus, depending upon the randomness of \textit{RS}, PDP ensures the highest level of security standard. From the usability perspective, PDP significantly raises the bar compared to ``take-a-tail" in terms of system interference and stress on memorability and makes  PDP highly practical approach to be used by common users. Most importantly, PDP reduces the storage overhead compared to all existing security approaches $-$ by storing a single information which is a huge benefit.

\section{Related work}
\label{relate}
The modern password cracking algorithm uses the concept of probabilistic context free grammars \cite{weir}. In \cite{guess}, Kelley et al. characterizes the vulnerability of the passwords under the same threat model \cite{weir} by considering different password-compositions policies. One of such weak password composition policy is ``basic8" in which users are instructed ``Password must have atleast 8 characters". One billion guess is sufficient to guess $40.3\%$ of such passwords. In \cite{graphical-processing-unit}, authors show that by using a single graphical processing unit, three billion guesses per second can be achievable to crack the hash functions like - MD5. Among the $70$ million yahoo users it has been observed that majority of the passwords are having little more than $20$ bits of effective entropy \cite{entropy} against an optimal attacker \cite{entropy2}.
The \textit{honeyword} scheme gives tremendous support to the conventional password scheme in terms of providing security and can be incorporated with the conventional password system. To the best of our knowledge, in $2006$ Fred Cohen has made the first contribution in this domain \cite{fakeaccount1}. There after many methodologies have been proposed in this direction. The idea has been deployed to many password related domains. Herley and Florencio \cite{online-brute} use this concept to protect online banking accounts from brute-force attack. Bojinov et al. propose the concept of ``Kamouflage" where real password of the user is stored along with the fake passwords but this does not include the concept of ``honeychecker" server \cite{kamouflage}. Later in \cite{juels}, authors introduce the concept of ``honeychecker" server to detect the password cracking mechanism. Recently Chakraborty and Mondal show how \textit{honeywords} can be used to detect shoulder surfing attack \cite{tag-honeypot}.

\section{Conclusion}
\label{conclusion}
Honeyword based techniques are getting popular as it provides several advantages over traditional password based schemes. However, the storage cost is one of the major overhead of honeyword based schemes.  
In this paper we have proposed a novel \textit{honeyword} generation approach which reduces the storage overhead and also it addresses majority of the drawbacks of existing \textit{honeyword} generation techniques. The only shortfall of PDP is, user has to remember an extra information in terms of \textit{RS}. In future we would like to analyse the possibility of developing a \textit{honeyword} generation technique without remembering any extra information by the users.

\bibliographystyle{abbrv}



%

\end{document}